\documentclass[twocolumn,showpacs,preprintnumbers,
amsmath,amssymb,aps,prl]{revtex4}

\usepackage{bm}
\usepackage{graphicx}

\bibstyle{apsrev}

\begin{document}

\title{Simultaneous Onset of Condensation of Molecules and Atoms\\
in an Attractive Fermi Gas of Atoms}
\author{Fuxiang Han, Minghao Lei, and E Wu}
\affiliation{Department of Physics,
Dalian University of Technology,
Dalian, Liaoning 116024,
China}
\date{\today}

\begin{abstract}
The self-consistent equations for the order parameters of Bose-Einstein 
condensation (BEC) of molecules and Bardeen-Cooper-Schrieffer (BCS) 
condensation of atoms in a Fermi gas of atoms with an attractive 
two-body interaction between atoms have been derived within the 
Hartree-Fock-Bogoliubov approximation from the path integral 
representation of the grand partition function. We have found that the 
order parameters for BEC and BCS are proportional to each other, which 
implies that BEC and BCS onsets simultaneously. We have also found that 
the common critical temperature of BEC and BCS increases as the average 
number of molecules increases and that the atom-molecule coupling 
enhances the common critical temperature.
\end{abstract}


\maketitle

The crossover regime~\cite{Zwierlein2003prl, Regal2004prl, Falco2004prl, 
Zwierlein2004prl, Partridge2005prl, Baym2005np, Chen2005pr, 
Greiner2005prl} of Bose-Einstein condensation (BEC) of molecules and 
Bardeen-Cooper-Schrieffer (BCS) condensation of atoms in a Fermi gas of 
atoms has recently been an active research field since it contains 
abundant information on the interplay of two kinds of quantum fluids 
(the Bose and Fermi quantum liquids) that exist in the nature. Progress 
in understanding of this crossover regime will certainly furnish 
significant implications and shed much light on our understanding of 
a plethora of macroscopic quantum phenomena including high temperature 
superconductivity. However, despite that great efforts have been put 
into it, the nature of condensation on either side of the crossover 
regime still remains an open question.

In this Letter, we report our results obtained in our study of 
Bose-Einstein condensation of molecules and BCS condensation of atoms on 
the BCS side in a Fermi gas of atoms with an attractive two-body 
interaction between atoms (an attractive Fermi gas of atoms). We will 
first derive a set of self-consistent equations for the order parameters 
of BEC and BCS in the Hartree-Fock-Bogoliubov approximation from the 
path integral representation of the grand partition function. Then, from 
the self-consistent equations we will draw the central conclusion 
in this Letter that BEC of molecules and BCS of atoms onset 
simultaneously in an attractive Fermi gas of atoms. We also calculate 
the common critical temperature of BEC and BCS and discuss its 
dependence on the fraction of atoms bound into molecules, on the 
two-body interaction between atoms, and on the atom-molecule coupling.

In the literature, the prime examples of Fermi gases of atoms are the 
gases of $^{40}$K and $^{6}$Li atoms, with the gas of $^{40}$K atoms 
most intensively studied so far. The magnetic-field Feshbach 
resonance~\cite{Stwalley1976prl, Tiesinga1993pra} is used to vary the 
sign and strength of the two-body interaction between atoms so that the 
BEC-BCS regime is conveniently accessed. In our present work, we are 
concerned with the BCS side in which the two-body interaction between 
atoms is attractive. On the BCS side, there exist an open channel and a 
closed channel, with bound states only in the closed channel. The 
bound-states in this case are referred to as unstable (unphysical) bound 
states. A resonance molecule refers to the entity formed by two atoms in 
such a bound state in real space. If the size of a bound state is small 
in real space, the bound state is then a conventional molecule. The size 
of a bound state can be varied by varying the strength of the two-body 
interaction between atoms through the Feshbach resonance. As its size 
increases, the bound state becomes better localized in momentum space 
and it can then be referred to as a preformed Cooper pair.

A Fermi gas of atoms with an attractive two-body interaction is 
described by the Hamiltonian~\cite{Holland2001prl}
\begin{eqnarray}
H &=& 2\Delta\nu\sum_{\bm{p}} a_{\bm{p}}^{\dagger}a_{\bm{p}}
+\sum_{\bm{k}\sigma}(\epsilon_{\bm{k}}-\mu)
c_{\bm{k}\sigma}^{\dagger}c_{\bm{k}\sigma}\nonumber\\
&&{}-\frac{|U|}{N}\sum_{\bm{k}\bm{k}^{\prime}\bm{p}}
c_{\bm{k}+\bm{p},\uparrow}^{\dagger}
c_{\bm{k}^{\prime}-\bm{p},\downarrow}^{\dagger}
c_{\bm{k}^{\prime}\downarrow}c_{\bm{k}\uparrow}\nonumber\\
&&{}+\frac{g}{\sqrt{N}}\sum_{\bm{p}\bm{k}}
\bigl(a_{\bm{p}}^{\dagger}c_{\bm{k}\uparrow}
c_{\bm{p}-\bm{k},\downarrow}+h.c.\bigr),
\label{Hamiltonian:U<0}
\end{eqnarray}
where $2\Delta\nu = 2(\nu-\mu)$ is the offset energy of a molecule 
relative to the chemical potential $\mu$, $a_{\bm{p}}^{\dagger}$ and 
$a_{\bm{p}}$ are creation and annihilation operators of a molecule of 
momentum $\bm{p}$ and energy $2\Delta\nu$, $c_{\bm{k}\sigma}^{\dagger}$ 
and $c_{\bm{k}\sigma}$ are creation and annihilation opperators of a 
fermionic atom of momentum $\bm{k}$, spin $\sigma$, and energy 
$\epsilon_{\bm{k}}=\hbar^2k^2/2m$, $U$ ($<0$) is the two-body 
interaction between atoms, $g$ is the effective atom-molecule coupling 
constant, and $N$ is the average number of single atoms.

Our starting point is the path integral representation~\cite{Negele1998, 
Zinn-Justin1996} of the grand partition function 
$Z=\mathrm{Tr}\,e^{-\beta H}$ with $H$ given in 
Eq.~\eqref{Hamiltonian:U<0}. To obtain the path integral representation 
of the grand partition function, we introduce the complex variables 
$\varphi_{\bm{p}}^{\ast}(\tau)$ and $\varphi_{\bm{p}}(\tau)$ for the 
creation and annihilation operators $a_{\bm{p}}^{\dagger}(\tau)$ and 
$a_{\bm{p}}(\tau)$ of molecules, and the Grassman variables 
$\psi_{\bm{k}\sigma}^{\ast}(\tau)$ and $\psi_{\bm{k}\sigma}(\tau)$ for 
the creation and annihilation operators 
$c_{\bm{k}\sigma}^{\dagger}(\tau)$ and $c_{\bm{k}\sigma}(\tau)$ of 
atoms. Replacing $a_{\bm{p}}^{\dagger}(\tau)$, $a_{\bm{p}}(\tau)$ by 
$\varphi_{\bm{p}}^{\ast}(\tau)$, $\varphi_{\bm{p}}(\tau)$ and 
$c_{\bm{k}\sigma}^{\dagger}(\tau)$, $c_{\bm{k}\sigma}(\tau)$ by 
$\psi_{\bm{k}\sigma}^{\ast}(\tau)$, $\psi_{\bm{k}\sigma}(\tau)$ in the 
Hamiltonian in Eq.~\eqref{Hamiltonian:U<0} so that the Hamiltonian 
becomes a function of the above-introduced complex and Grassman 
variables and is denoted by $H(\varphi, \psi)$, and then substituting 
$H(\varphi,\psi)$ into\begin{eqnarray}
Z &=& \int D[\varphi^{\ast},\varphi;\psi^{\ast},\psi]_{\tau}\;
\exp\biggl\{-\int_{0}^{\beta}d\tau\;\Bigl[
\varphi_{\bm{p}}^{\ast}(\tau)
\frac{\partial}{\partial\tau}\varphi_{\bm{p}}(\tau)\nonumber\\
&&{}+\psi_{\bm{k}\sigma}^{\ast}(\tau)
\frac{\partial}{\partial\tau}\psi_{\bm{k}\sigma}(\tau)
-H(\varphi,\psi)\Bigr]\biggr\},
\label{path integral:general}
\end{eqnarray}
we obtain the integral representation of the grand partition function. 
The integration measure in Eq.~\eqref{path integral:general} is given by
\begin{eqnarray}
D[\varphi^{\ast},\varphi;\psi^{\ast},\psi]_{\tau}
&=& \lim_{M\rightarrow\infty}\prod_{j=1}^{M}\prod_{\bm{k}\sigma}
d\psi_{\bm{k}\sigma}(j\Delta\tau)
d\psi_{\bm{k}\sigma}^{\ast}(j\Delta\tau)\nonumber\\
&&{}\times
\prod_{\bm{p}}\frac{d\varphi_{\bm{p}}^{\ast}(j\Delta\tau)
d\varphi_{\bm{p}}(j\Delta\tau)}{2\pi i}
\end{eqnarray}
with $\Delta\tau=\beta/M$.

The complex variable $\varphi_{\bm{p}}(\tau)$ introduced above acts as 
the order parameter of Bose-Einstein condensation of molecules. To 
proceed further, we use Fourier transformations of 
$\varphi_{\bm{p}}(\tau)$ and $\psi_{\bm{k}\sigma}(\tau)$ given by 
$\varphi_{\bm{p}}(\tau) = \sum_{i\omega_m}e^{-i\omega_{m}\tau} 
\varphi_{p}$ and $\psi_{\bm{k}\sigma}(\tau) = 
\sum_{i\omega_n}e^{-i\omega_{n}\tau}\psi_{k\sigma}$ 
[where $i\omega_m=i2m\pi/\beta$ and $i\omega_n=i(2n+1)\pi/\beta$, with 
$m$ and $n$ being integers, are Matsubara imaginary frequencies for 
bosons and fermions, respectively, and $p=(\bm{p}, i\omega_m)$, 
$k=(\bm{k}, i\omega_n)$] to rewrite the path integral representation in 
imaginary frequency space. For the purpose of decoupling the two-body 
interaction between atoms through the Hubbard-Stratonovich 
transformation~\cite{Hubbard1972pl, Stratonovich1958spd}, we rewrite it 
as
\begin{eqnarray}
\lefteqn{\frac{1}{N}\sum_{kk^{\prime}p}\psi_{k+p,\uparrow}^{\ast}
\psi_{k^{\prime}-p,\downarrow}^{\ast}\psi_{k^{\prime}\downarrow}
\psi_{k\uparrow}}\nonumber\\
&=& \sum_{p}\biggl[\frac{1}{\sqrt{N}}\sum_{k}
\psi_{p-k\downarrow}\psi_{k\uparrow}\biggr]^{\ast}
\biggl[\frac{1}{\sqrt{N}}\sum_{k}
\psi_{p-k\downarrow}\psi_{k\uparrow}\biggr].\;\quad
\end{eqnarray}
The field introduced in implementing the Hubbard-Stratonovich 
transformation is denoted by $\Delta_{p}$ and it corresponds to the 
order parameter of BCS condensation of atoms~\cite{Negele1998}. For 
discussion of Bose-Einstein condensation of molecules and BCS 
condensation of atoms, we need retain only zero bosonic 
momentum-frequency terms in the atom-molecule coupling and in the 
decoupled terms. Then all nonzero momentum-frequency order parameters 
can be integrated out, with only the order parameters of zero 
momentum and frequency [they are now simply denoted by $\varphi$ and 
$\Delta$, respectively] remaining in the path integral representation of 
the grand partition function.

The next algebraic manipulation to perform is the integration over 
the Grassman variables. To integrate out the Grassman variables, we 
first bring the terms containing the Grassman variables as a whole into 
a diagonal form. This manipulation is similar to the diagonalization of 
the Hamiltonian of a system of fermions~\cite{Blaizot1986}. We found 
that it is convenient to introduce a column matrix of two Grassman 
variables and to write the concerned terms in a matrix form. Then we 
diagonalize the matrix appearing in the matrix form of the concerned 
terms. The transformation matrix that transforms the to-be-diagonalized 
matrix into a diagonal matrix is composed of the orthonormal 
eigenvectors of the to-be-diagonalized matrix. The new Grassman 
variables in terms of which the concerned terms are expressed in a 
diagonal form can be easily obtained by using the transformation matrix. 
It can be directly verified that the Jacobian of the change of variables 
from the old Grassman variables to the new Grassman variables is unity.

The final simplification we can do to the path integral representation 
of the grand partition function is to perform the summations over the 
imaginary frequencies $i\omega_m$ and $i\omega_n$. These summations can 
be easily performed by considering proper contour 
integrals~\cite{Abrikosov1975, Mahan1990}. With the summations over 
$i\omega_m$ and $i\omega_n$ done, we obtain
\begin{equation}
Z = N^2\beta |U|\int \frac{d\varphi^{\ast}d\varphi}{2\pi i}
\frac{d\Delta^{\ast}d\Delta}{2\pi i}\;e^{-S},
\end{equation}
where
\begin{eqnarray}
S &=& \sum_{\bm{p}\neq 0}
\ln\frac{\sinh(\beta\Delta\nu)}{\beta\Delta\nu}
-2\sum_{\bm{k}}\ln\biggl[2\cosh\biggl(\frac{
\beta E_{\bm{k}}}{2}\biggr)\biggr]\nonumber\\
&&{}+2N\beta\Delta\nu|\varphi|^{2}
+N\beta |U||\Delta|^{2}
\label{action:general}
\end{eqnarray}
with
\begin{equation}
E_{\bm{k}} = \left(\xi_{\bm{k}}^2+|g\varphi
+|U|\Delta|^2\right)^{1/2}.
\end{equation}

Notice that the prefactor $N^2\beta |U|$ in Eq.~\eqref{action:general} 
arises from the change of variables $\varphi \rightarrow \sqrt{N} 
\varphi$ and $\Delta \rightarrow \sqrt{N} \beta |U| \Delta$ before 
performing the summations over $i\omega_m$ and $i\omega_n$. The purpose 
of this change of variables is to make the order parameters $\varphi$ 
and $\Delta$ intensive.

An approximate explicit expression of $Z$ can be obtained by making use 
of the steepest-descent method~\cite{Courant1962} in which $S$ is 
expanded as a Taylor series in $\varphi$ and $\Delta$ about the 
stationary point of $S$. As an approximation, this series is terminated 
at the second order. The integration over $\varphi$ and $\Delta$ can 
then be performed and an approximate explicit expression of $Z$ is 
obtained. The stationary value of $S$ is  obtained by evaluating its 
value at the stationary point determined by the conditions that 
$\partial S/\partial\varphi^{\ast} = 0$ and $\partial 
S/\partial\Delta^{\ast} = 0$ and their complex conjugates. From the 
stationary conditions, we obtain the self-consistent equations for 
$\varphi$ and $\Delta$
\begin{subequations}
\label{sces}
\begin{eqnarray}
&& \varphi=\frac{g}{4N\Delta\nu}\bigl(g\varphi+|U|\Delta\bigr)
\sum_{\bm{k}}\frac{1}{E_{\bm{k}}}
\tanh\biggl(\frac{\beta E_{\bm{k}}}{2}\biggr),\quad
\label{sce:varphi}\\
&& \Delta= \frac{1}{2N} \bigl(g\varphi+|U|\Delta\bigr)
\sum_{\bm{k}}\frac{1}{E_{\bm{k}}}\tanh\biggl(\frac{\beta 
E_{\bm{k}}}{2}\biggr).
\label{sce:Delta}
\end{eqnarray}
\end{subequations}
The self-consistent equations for the complex conjugates of $\varphi$ 
and $\Delta$ can be obtained simply by taking the complex conjugation 
of the above two equations.

From Eqs.~\eqref{sces}, we immediately see that $\varphi$ and $\Delta$ 
are proportional to each other
\begin{equation}
\varphi = \frac{g}{2\Delta\nu}\Delta.
\label{varphi-delta relation}
\end{equation}
The proportionality between $\varphi$ and $\Delta$ implies that they 
become nonzero at the same temperature as the temperature is lowered 
from a value higher than the critical temperature. In other words, 
Bose-Einstein condensation of molecules and BCS condensation of atoms 
occur simultaneously. This result is of great significance in that it 
reveals to us that both molecules and atoms condense in an attractive 
Fermi gas of atoms and furthermore that their condensation onsets at a 
common temperature. Physically, the simultaneity of condensation of 
molecules and atoms is due to the presence of the atom-molecule 
coupling.

We now turn to the evaluation of the common critical temperature of 
condensation of molecules and atoms. The common critical temperature 
$T_c$ can be determined from any one of the two self-consistent 
equations in Eqs.~\eqref{sces} in conjunction with the relation in 
Eq.~\eqref{varphi-delta relation}. Explicitly, we have
\begin{equation}
1 = \biggl(\frac{g^2}{2\Delta\nu}+|U|\biggr)
\frac{1}{2N}\sum_{\bm{k}}\frac{1}{\xi_{\bm{k}}}
\tanh\biggl(\frac{\xi_{\bm{k}}}{2k_BT_c}\biggr).
\label{sce:Tc}
\end{equation}

Notice that, since the chemical potential appears in $\Delta\nu = 
\nu-\mu$, the chemical potential $\mu$ has to be determined 
self-consistently. To determine $\mu$, we set up a self-consistent 
equation for it by using the relation $2M+N=-\partial F/\partial \mu$ in 
which $F$ is the Helmholz free energy and $M$ is the average number of 
molecules. This relation can be easily verified by using $F=-\beta^{-1} 
\ln Z$ and $Z = \mathrm{Tr}\,e^{-\beta H}$ with $H$ given in 
Eq.~\eqref{Hamiltonian:U<0}. With $Z$ expressed as $Z=e^{-S}$, the above 
relation can be written as $2M+N=-\beta^{-1} \partial S/ \partial \mu$. 
In the Hartree-Fock-Bogoliubov approximation, it is satisfactory to 
approximate $S$ with its stationary value. With the utilization of such 
an approximation to $S$, we obtain the following self-consistent 
equation for the chemical potential $\mu_c$ at $T=T_c$

\begin{equation}
1 = \eta_{m}\biggl(
\frac{1}{e^{2\beta_c\Delta\nu}-1}
-\frac{1}{2\beta_c\Delta\nu}\biggr)
+\frac{2\eta_{a}}{N}\sum_{\bm{k}}\frac{1}{e^{\beta_c \xi_{\bm{k}}}+1},
\label{sce:muc}
\end{equation}
where $\beta_c=1/k_BT_c$, $\eta_{m}$ is the ratio of the number of atoms 
bound into molecules to the total number of atoms, $\eta_{m} = 
2M/(2M+N)$, and $\eta_{a}$ is the ratio of the number of single atoms to 
the total number of atoms, $\eta_{a} = N/(2M+N)$. Notice that, in 
deriving the self-consistent equation for $\mu_c$, the zero-point 
contributions must be omitted~\cite{Andersen2004rmp}.

Before we proceed to solve for the common critical temperature $T_c$ 
from Eqs.~\eqref{sce:Tc} and~\eqref{sce:muc}, we first specify values 
for several parameters. Firstly, an energy cutoff is required in 
Eq.~\eqref{sce:Tc}. This cutoff indicates the energy range of atoms that 
participate in pairing. We take this energy cutoff to be the Fermi 
energy $E_F$ since all atoms are allowed to participate in pairing . For 
a typical density of $n = 10^{20}\;\mathrm{m}^{-3}$~\cite{Regal2004prl, 
Holland2001prl}, $E_F \approx 1.25\;\mu\mathrm{K}$ for $^{40}$K atoms. 
Secondly, the range of the atom-molecule coupling $g$ we will consider 
is between $0.1 E_F$ and $E_F$. Thirdly, for the two-body interaction 
between atoms, we use $U = -0.2 E_F$. Lastly, for the molecular offset 
energy, we use $\nu = E_F$. The above-quoted values of parameters are in 
consistency with the experiment~\cite{Regal2004prl} and with the 
previous theoretical analysis~\cite{Holland2001prl}.

\begin{figure}[htb]
\includegraphics{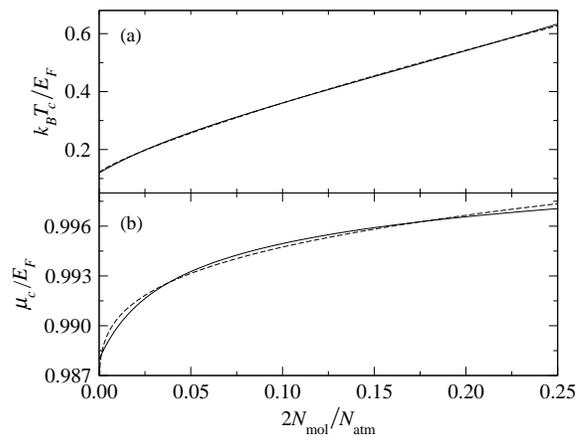}
\caption{\label{fig:1} The dependence of the common 
critical temperature $T_c$ [(a)] and the chemical potential $\mu_c$ at 
$T_c$ [(b)] on the fraction of atoms bound into molecules for $g = 
0.1E_F$, $U=-0.2E_F$, and $\nu=E_F$. The dashed lines are power-law fits 
discussed in the text.}
\end{figure}

Shown in Fig.~\ref{fig:1}(a) [the solid line] is the 
dependence of the common critical temperature $T_c$ on the fraction 
$\eta_m$ of atoms bound into molecules, with the dependence on $\eta_m$ 
of the chemical potential $\mu_c$ at $T_c$ shown in Fig.~\ref{fig:1}(b). 
It is seen that, as $\eta_m$ increases, $T_c$ increases almost linearly, 
with $T_c \approx 0.12E_F$ at $\eta_m = 0$ and $T_c \approx 0.63 E_F$ at 
$\eta_m = 0.25$. We thus conclude that the presence of more molecules 
effectively raises the common critical temperature of condensation. The 
dependence of $T_c$ on $\eta_m$ can be well fitted with the power law 
$k_BT_c/E_F = a + b \eta_m^{\gamma}$ [shown in Fig.~\ref{fig:1}(a) as a 
dashed-line] with $a \approx 0.120$, $b \approx 1.569$, and $\gamma 
\approx 0.815$. From Fig.~\ref{fig:1}(b), we can see that, as $\eta_m$ 
increases, the value of the chemical potential $\mu_c$ at $T_c$ also 
increases, with the increase slowing down at large values of $\eta_{m}$. 
The dashed-line in Fig.~\ref{fig:1}(b) is the fit to the 
power law $k_BT_c/E_F = c + d \eta_m^{\delta}$ with $c = 0.984$, $d = 
1.824\times 10^{-2}$, and $\delta = 0.233$.

\begin{figure}[htb]
\includegraphics{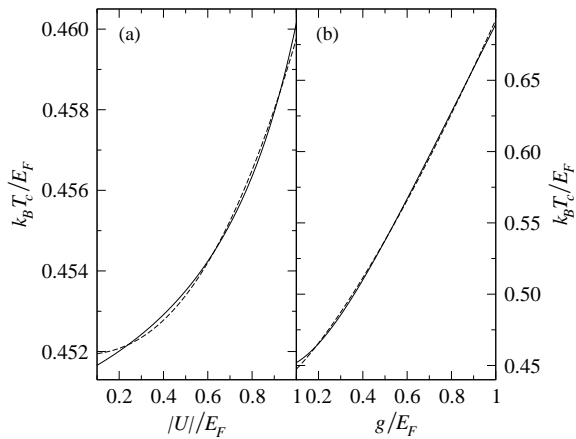}
\caption{\label{fig:2} The effects of the two-body interaction [(a)] and 
the atom-molecule coupling [(b)] on the common critical temperature 
$T_c$. The dashed lines are the power-law fits discussed in the text.}
\end{figure}

Both the two-body interaction between atoms and the atom-molecule 
coupling have significant influences on the common critical temperature 
$T_c$ as shown in Fig.~\ref{fig:2} for $\eta_m = 0.15$ and $\nu 
= E_F$. At these values of $\eta_m$ and $\nu$, the values of $T_c/E_F$ 
fall in the vicinity of $0.5E_F$ for values of $|U|/E_F$ and $g/E_F$ 
within the range of $0.1 \sim 1.0$, which is in consistency with the 
experiment~\cite{Regal2004prl}. As $|U|$ or $g$ increases, $T_c$ 
increases monotonically. It has been found that both the 
$T_c$-versus-$|U|$ and $T_c$-versus-$g$ curves can be well fitted with a 
power law $k_BT_c/E_F = \alpha + \beta x^{\rho}$. For the 
$T_c$-versus-$|U|$ curve, $x=|U|/E_F$, $\alpha = 0.452$, $\beta = 
7.85\times 10^{-3}$, and $\rho = 2.410$; for the $T_c$-versus-$g$ 
curve, $x=g/E_F$, $\alpha=0.435$, $\beta = 0.257$, and $\rho=1.327$. 
These fits are shown in Figs.~\ref{fig:2}(a) and~\ref{fig:2}(b) as 
dashed-lines. From Fig.~\ref{fig:2}, it can be seen that the influence 
on $T_c$ of the atom-molecule coupling is stronger than that of the 
two-body interaction between atoms. The reason for this behavior is the 
fact that $g$ appears in the quadratic form in the effective coupling 
constant $(g^2/2\Delta\nu+|U|)$ [the first factor on the right hand side 
of Eq.~\eqref{sce:Tc}], while $|U|$ appears only in the linear form.

In encapsulation, in terms of the self-consistent equations for the 
order parameters derived from the path integral representation of the 
grand partition function, we have studied Bose-Einstein condensation of 
molecules and BCS condensation of atoms in an attractive Fermi gas of 
atoms as realized in experiments~\cite{Regal2003prl, Regal2004prl, 
Greiner2005prl} by making use of the magnetic-field resonance. We have 
found that molecules and atoms start to condense at the same critical 
temperature, that the common critical temperature increases as the 
number of molecules increases, and that the atom-molecule coupling 
raises the common critical temperature.

Our findings have a number of important implications both to the 
BEC-BCS crossover in particular and to superfluidity (superconductivity) 
in general. The fact that both molecules and atoms condense and that 
they have the very same critical temperature resolves the controversy 
over the nature of condensation on the BCS side of the BEC-BCS crossover 
regime and manifests that the condensate in an attractive Fermi gas of 
atoms is actually composed of a Bose-Einstein condensate and a BCS 
condensate. The increase of the common critical temperature with the 
fraction of atoms bound into molecules indicates that high temperature 
(possibly room temperature) superfluidity (superconductivity) is 
likely to be achieved in a system in which Bose-Einstein condensation of 
bosons and BCS condensation of fermions can occur simultaneously. The 
enhancement of the common critical temperature by the atom-molecule 
coupling suggests that high temperature superfluidity 
(superconductivity) is likely to be discovered in a system with a strong 
boson-fermion interaction in addition to that both bosons and fermions 
in the system can undergo, respectively, Bose-Einstein and BCS 
condensations.

\bibliography{simul}

\end{document}